\documentclass[twocolumn,showpacs,showkeys,preprintnumbers,amsmath,amssymb,nofootinbib]{revtex4}
\usepackage{graphicx}
\usepackage{dcolumn}
\usepackage{bm}
\begin{document}
\title{\bf Virtual Photon Emission from Quark-Gluon Plasma }
\author{S.  V.  Suryanarayana}
\email{snarayan@barc.gov.in}
\email{suryanarayan7@yahoo.com}
\affiliation{  Nuclear Physics Division, Bhabha Atomic Research Centre, Trombay, Mumbai
400 085, India}
\begin{abstract}{We  recently proposed an empirical approach for the  Landau-Pomeranchuk-Migdal  (LPM) effects 
in  photon emission from the quark gluon plasma as a function of photon mass. This approach was
based on Generalized Emission Functions (GEF) for photon emission, derived  at a  fixed temperature and strong coupling constant. 
In the present  work,  we have extended the LPM calculations for several temperatures and 
strong coupling strengths. The integral equations for  (${\bf   \tilde{f}(\tilde{p}_\perp  )}$)  and   ($\tilde{g}({\bf \tilde{p}_\perp } )$) 
are solved by the  iterations method for  the variable set \{$p_0,q_0,Q^2,T,\alpha_s$\}, considering 
bremsstrahlung and  $\bf aws$ processes. We generalize the dynamical scaling variables,  $x_T$,  $x_L$, 
for bremsstrahlung and {\bf aws} processes which are  now functions  of variables $p_0,q_0,Q^2,T,\alpha_s$.
The GEF  introduced earlier, $g^b_T$,   $g^a_T$,   $g^b_L$,   $g^a_L$,  are also generalized for any 
temperatures and coupling strengths.  From this,  the  imaginary  part   of  the  photon  polarization tensor 
as a function of photon mass and energy can be calculated as a one dimensional integral over these GEF and parton 
distribution functions in the plasma.  However, for phenomenological studies of experimental 
data, one needs a simple empirical formula 
without involving parton momentum integrations. Therefore,  we present a phenomenological  formula for imaginary photon 
polarization tensor  as a function of  \{$q_0,Q^2,T,\alpha_s$\} that includes  bremsstrahlung and $\bf aws$ mechanisms along 
with LPM effects.
}\end{abstract}
\maketitle
\par
\noindent
In this  work, we present a study of Landau-Pomeranchuk-Migdal effects \cite{landau1,migdal} (LPM) 
in virtual  photon emission from thermalized quark gluon plasma (QGP).
The LPM effects  on real photon emission from QGP have been reported \cite{arnold1,arnold2} 
and an empirical approach in \cite{svsprc}.
For the case of virtual photon emission in QGP, the processes that contribute 
at  $\alpha\alpha_s$  order \cite{alther} and the  higher order corrections  \cite{thoma} 
and LPM effects \cite{lpmdilep} were well studied. In hard thermal loops (HTL) \cite{braaten} method 
these processes occur at the  one  loop, two lop and higher loop  levels represented by ladder diagrams.
In the photon emission  calculations, the quantity of interest is the the imaginary part of  photon retarded 
polarization tensor ($\Im{\Pi^\mu}_{R\mu}$). The dilepton emission rates  are estimated in terms of  
this $\Im{\Pi^\mu}_{R\mu}$,  Bose-Einstein factor and $Q^2$  as  given by Eq.\ref{dileprate}. 
The  $\Im{\Pi^\mu}_{R\mu}$ including LPM effects is determined  in terms of 
a transverse  function    ${\bf {f}({{p}}_\perp)}$ and a longitudinal part ${{g}({{p}}_\perp)}$, as 
given by Eq.\ref{impolar} \cite{lpmdilep}.
\par
For  the  case  of  virtual  photon  emission  having small virtuality,  the
transverse vector function  ${\bf f({p}_\perp})$ is  determined  by  the  AMY equation (Eq.\ref{agmz-t-tilde})
and the longitudinal function by  AGMZ equation (Eq.\ref{agmz-l-tilde}) \cite{lpmdilep}. The energy    transfer  function   
$\delta    {E}({\bf{p}_\perp},p_0,q_0,Q^2,T,\alpha_s)$  is given in  Eq.\ref{deltae}.  The tilde represents quantities in 
units of Debye mass, for details see \cite{svsarxiv06}.
\begin{eqnarray}
\frac{dN_{\ell^+\ell^-}}{d^4xd^4Q} &=& \frac{\alpha_{EM}}{12\pi^4Q^2(e^{q0/T}-1)} \Im\Pi_{R\mu}^\mu(Q)  \label{dileprate} \\
\Im{\Pi^\mu}_{R\mu} &=& \frac{e^2N_c}{2\pi} \int_{-\infty}^\infty  dp_0 [n_F(r_0)-n_F(p_0)] \otimes \nonumber \\
&& \int \frac{d^2{\bf {p}_\perp}}{(2\pi)^2}\left[ \frac{p_0^2+r_0^2}{2(p_0r_0)^2} \Re{\bf {p}_\perp.{f}({p}_\perp)} + \right. \nonumber \\
&& \left. \frac{1}{\sqrt{\left|p_0r_0\right|}}\frac{Q^2}{q^2} \Re {g}({\bf{ p}_\perp})\right] 
\label{impolar}
\end{eqnarray}
\begin{eqnarray}
2{\bf \tilde{p}_\perp}&=& i \widetilde{\delta E}({\bf \tilde{p}_\perp},p_0,q_0,Q^2)
{\bf \tilde{f}}({\bf \tilde{p}_\perp})  \nonumber \\
&& + \int\frac{d^2{\bf \tilde{\bf\ell}_\perp}}{(2\pi)^2} \tilde{C}({\bf \tilde{\ell}_\perp})
\left[{\bf \tilde{f}}({\bf \tilde{p}_\perp}) -{\bf \tilde{f}}({\bf \tilde{p}_\perp+\tilde{\bf\ell}_\perp})\right] \label{agmz-t-tilde}\\
2\sqrt{\frac{|p_0r_0|}{m_D^2}}&=& i\widetilde{\delta E}(({\bf \tilde{p}_\perp},p_0,q_0,Q^2)
{\tilde{g}}({\bf \tilde{p}_\perp}) + \nonumber \\
&& \int\frac{d^2{\bf \tilde{\ell}_\perp}}{(2\pi)^2}\tilde{C}({\bf \tilde{\ell}_\perp}) \left[{\tilde{g}}({\bf \tilde{p}_\perp}) -{\tilde{g}}({\bf \tilde{p}_\perp+\tilde{\bf\ell}_\perp})\right] ~~~~~
\label{agmz-l-tilde} \\
\widetilde{\delta E}&=& \frac{q_0T}{2p_0(q_0+p_0)}\left[\tilde{p}_\perp^2+\kappa_{\mbox{\small{eff}}}\right] 
\label{deltae}
\end{eqnarray}
\section{Generalized Emission Functions for photon emission}
\begin{figure}
\hspace{-1.05cm}
\includegraphics[height=14.cm,width=9.5cm]{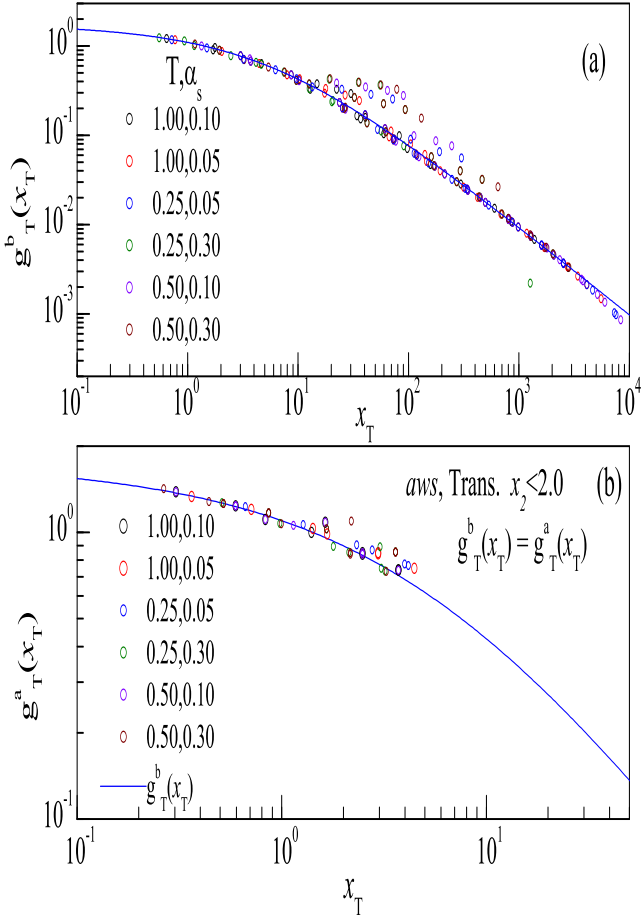}
\caption{  (a)~ The  dimensionless  emission  function  $g^b_T(x)$  versus dynamical
variable $x_T$ defined in Eq.\ref{xt}.  Six cases of  temperature and coupling constant values 
considered are mentioned in figure labels in different colored symbols.
The symbols represent the integrated values of
${\bf p_\perp}$ distributions as a function  of $\{p_0,q_0,Q^2,T,\alpha_s\}$ values.
These are transformed by  $c^b_T$  coefficient function given in  Eq.\ref{cbt}. 
Essentially,  various symbols merge and can not be distinguished. 
The solid curve is an empirical fit given by Eq.\ref{btgxemp}.
(b)~The  dimensionless  emission  function  $g^a_T(x)$  versus dynamical
variable $x_T$ for $x_2<2.0$.  The transformation coefficients $c^a_T$
and  empirical fit are given by Eqs.\ref{cat},\ref{atgxemp}.}
\label{btatgx}
 \end{figure}

\begin{figure}
\hspace{-1.05cm}
\includegraphics[height=10.cm,width=9.5cm]{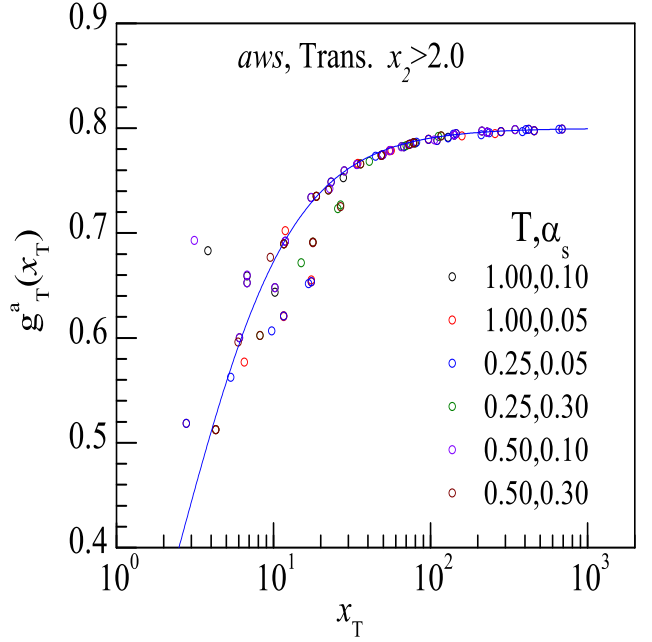}
\caption{  The  dimensionless  emission  function  $g^a_T(x)$  versus dynamical
variable $x_T$.   The symbols are as in figure \ref{btatgx}. Six different temperature and coupling 
constant values considered are mentioned in figure labels. The required 
$c^a_T$  coefficient function given in  Eq.\ref{cath}. 
The solid curve is an empirical fit given by Eq.\ref{btgxemp}.}
\label{athgx}
\end{figure}

\begin{figure}
\hspace{-1.05cm}
\includegraphics[height=16.cm,width=9.5cm]{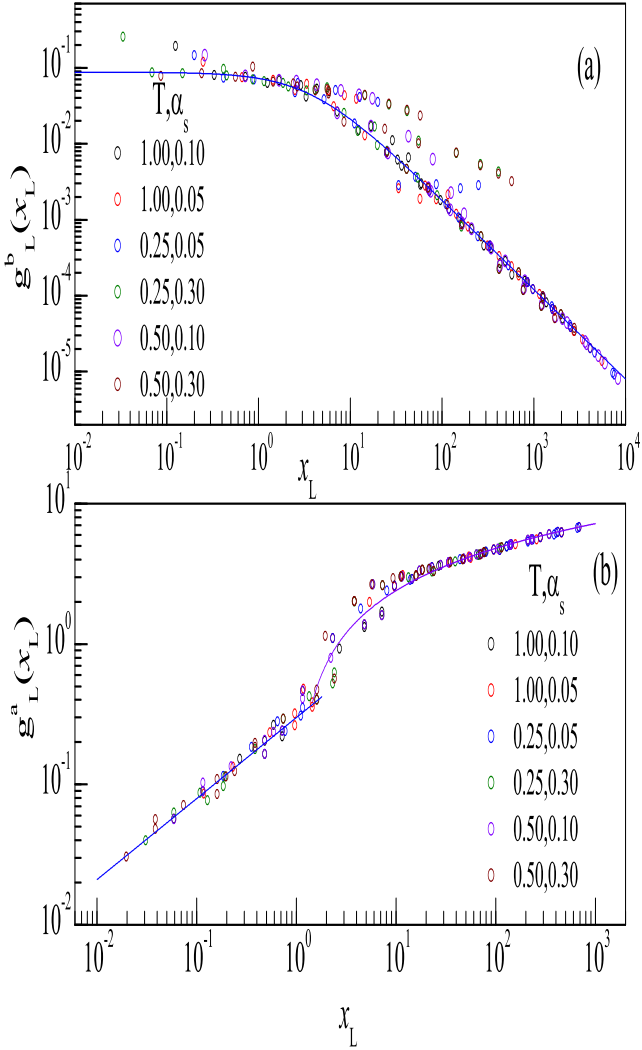}
\caption{  (a)~ The  dimensionless  emission  function  $g^b_L(x)$  versus dynamical
variable $x_L$ defined in Eq.\ref{xl}.  The symbols represent the integrated values of
${\bf p_\perp}$ distributions as a function of $\{p_0,q_0,Q^2,T,\alpha_s\}$ values.
These are transformed by  $c^b_L$  coefficient function given in  Eq.\ref{cbl}. 
The solid curve is an empirical fit given by Eq.\ref{blgxemp}. The temperature and coupling constant values 
are mentioned in figure labels in different colors.
(b)~The  dimensionless  emission  function  $g^a_L(x)$  versus dynamical
variable $x_L$.  The transformation coefficients $c^a_L$
and  empirical fit are given by Eqs.\ref{cal},\ref{algxemp}.}
\label{blalgx}
 \end{figure}

In the present work, we  solved these   Eqs.(\ref{agmz-t-tilde},\ref{agmz-l-tilde})  by iterations method 
at a fixed photon energy of ${q_0}$/T=50.
Alternatively, these equations can also be solved by variational approach \cite{svsvar}.
In the following calculations, we have used  two flavors and three colors. 
Using the iterations method,  we obtained ${\bf p_\perp.f(p_\perp)}, g({\bf p_\perp})$ distributions
for different $p_0,q_0,Q^2$,  plasma temperatures (T=1.0, 0.50, 0.25GeV) and strong coupling constants ($\alpha_s$=0.30, 0.10, 0.05).
We integrate these ${\bf p_\perp.f(p_\perp)}, g({\bf p_\perp})$ distributions to derive $I^{b,a}_{T,L}$ 
as defined in the Eqs.\ref{itdef},\ref{ildef}.
The superscripts $b,a$  in these equations represent bremsstrahlung or $\bf aws$ processes depending on the $p_0$ value used.
The subscripts $T,L$ represent contributions from transverse ($\bf f(p_\perp)$) or longitudinal $(g(\bf p_\perp))$ parts.
$I^{b,a}_{T,L}$ are the quantities required for calculating imaginary part of polarization tensor (see Eq.\ref{impolar}). 
Therefore, in the following,  we empiricize these $I^{b,a}_{T,L}$.
\begin{eqnarray}
I^{b,a}_{T}&=&\int \frac{\bf d^2\tilde{p}_\perp}{(2\pi)^2}  {\bf \tilde{p}_\perp \cdot\Re\tilde{f}(\tilde{p}_\perp)} \label{itdef} \\
I^{b,a}_{L}&=&\int \frac{\bf d^2\tilde{p}_\perp}{(2\pi)^2}  \Re\tilde{g}({\bf \tilde{p}_\perp)} \label{ildef} \\
x_0&=&\frac{|(p_0+q_0)p_0|}{q_0T  } ~;~x_3 = \frac{q_0T(\alpha_s/0.3)}{Q^2} \label{x0x3} \\
x_1 &=&x_0\frac{M_\infty^2}{m_D^2}  \label{x1}\\
x_2&=&x_0\frac{Q^2}{q_0T(\alpha_s/0.3)} \label{x2} \\
x_T&=&x_1+x_2   \label{xt}  \\
x_L&=&x_2      \label{xl}    \\
g^{b,a}_{T,L}(x_{T,L})&=&I^{b,a}_{T,L}(x_{T,L})c^{b,a}_{T,L}  \label{gbatl} 
\end{eqnarray}
In the remaining part of this work, we adopt the formulae and results of \cite{svsarxiv06} presented
at  fixed T=1GeV, $\alpha_s=0.30$, by suitably redefining the quantities for all temperatures and strong coupling constants.
In Eqs.\ref{x0x3},\ref{x1},\ref{x2} we define four dimensionless variables. The factor $\alpha_s/0.3$ in above  equations is required to match the definitions in present work with those of  \cite{svsarxiv06}.
The variable $x_1$ is the real photon dynamical variable  \cite{svsprc}.  
For virtual photon emission from QGP, we define two more variables,  
$x_{T,L}$  given in Eqs.\ref{xt},\ref{xl}.
 $I^{b,a}_{T,L}$ are in general functions of \{$p_0,q_0,Q^2,T,\alpha_s$\} 
and when plotted versus any of these $p_0,q_0,Q^2$, they do not exhibit any simple trends. 
Following \cite{svsarxiv06}, we define the generalized emission functions (GEF) $g^{b,a}_{T,L}$ in Eq.\ref{gbatl}. 
The  GEF are functions of only $x_{T,L}$  variables. 
These  GEF ($g^{b,a}_{T,L}$)  are obtained from corresponding $I^{b,a}_{T,L}$ values by multiplying with
$c^{b,a}_{T,L}$ coefficient functions given in Eqs.\ref{cbt}-\ref{cal}.
The variable $x$ in Eqs.\ref{btgxemp}-\ref{algxemp} is  $x_T$  for transverse part and  $x_{L}$ for longitudinal parts.
The quantities $x_{T,L}$ and  $c^{b,a}_{T,L}$  in Eqs.\ref{xt}-\ref{cal} are found by   search 
for  dynamical variables hidden in the solutions of AMY and AGMZ equations.
\begin{eqnarray}
c^b_T&=&\frac{1}{x_1^2}    \label{cbt}  \\
c^a_T&=&\frac{1}{x_1x_2} \label{cat}  \\
c^a_T&=&\frac{1}{x_1^2}\frac{x_3}{1+x_3}   ~~~\mbox{for}~~~ x_2<2.0   \label{catl}  \\
c^b_L&=&\frac{Q^2}{T^2(\alpha_s/0.3)}\left(\frac{T^2}{p_0(p_0+q_0)}\right)^{\frac{3}{2}} \otimes \nonumber \\
          & & \frac{\left(1.5+x_3^{0.75}\right)}{x_2^{1/3}}\sqrt{\frac{\alpha_s}{0.3}} \label{cbl}     \\
c^a_L&=&\frac{x_2^{0.10} }{x_1^{1.40}\sqrt{q_0/T}\left(1+\sqrt{x_3}\right)}\sqrt{\frac{\alpha_s}{0.3}}\label{cal}
\end{eqnarray}
\begin{eqnarray}
g^b_T(x)&=&\frac{10.0}{5.6+2.5\sqrt{x}+x}  \label{btgxemp} \\
g^a_T(x)&=&\frac{0.80}{(1+3/x^{1.2})}    \label{atgxemp} \\
g^a_T(x)&=&g^b_T(x) ~~~\mbox{for}~~~ x_2<2.0     \label{atlgxemp} \\
g^b_L(x)&=&\frac{0.0876}{1+\left(\frac{x}{3.7727}\right)^{1.18}}  \label{blgxemp} \\
g^a_L(x)&=&0.299803 x^{0.5772}   \label{algxemp} \\
g^a_L(x)&=&1.04344\ln{(x)} ~~~\mbox{for}~ x>1.45 \nonumber
\end{eqnarray}
Figure \ref{btatgx} shows the results for  GEF for bremsstrahlung (Fig.\ref{btatgx}(a)).
The calculations are for a fixed photon energy  ($q_0$/T=50.) 
but include six different cases of temperatures and coupling strengths mentioned 
in  figure labels. The solid curve in (a) is the empirical fit to this emission function, given by Eq.\ref{btgxemp}
\footnote{This fit given in Eq.\ref{btgxemp} is an improvement over the  result reported in \cite{svsarxiv06}.}.
The required  $c^b_T$ coefficient function is given in Eq.\ref{cbt}. 
 It has been observed that for low  Q$^2$,  {\it i.e.,} $x_2<2.0$,  transverse part of $ \bf aws$ process
behaves similar to the transverse bremsstrahlung function. Therefore, we transform the low 
 Q$^2$  transverse part of $\bf aws$ process as given by Eq.\ref {catl}. The resulting emission function is shown in  Fig. 1(b).
The solid curve is given in Eq.\ref{atlgxemp}, which is same as solid curve in Fig. 1(a).\\
The emission function for high Q$^2$ values ($x_2>2.0$) for transverse part of $\bf aws$ process is shown in Figure \ref{athgx}.
The $c^a_T$  and the emission function are given in Eqs.\ref{cat},\ref{atgxemp}. Similarly, Figures (\ref{blalgx}(a,b)
show the longitudinal components of GEF for bremsstrahlung (Fig.(a)) and $\bf aws$ (Fig.(b)). 
The  coeffiicient functions and GEF are given in Eqs.\ref{cbl},\ref{blgxemp},\ref{cal},\ref{algxemp}. 
These transformation functions are very complex.\footnote{The Eqs.\ref{cal},\ref{algxemp} are slightly 
different from the corresponding equations  presented in \cite{svsarxiv06}.}.
\section{GEF and photon retarded polarization tensor}
\begin{figure}
\hspace{-1.05cm}
\includegraphics[height=12.cm,width=9.5cm]{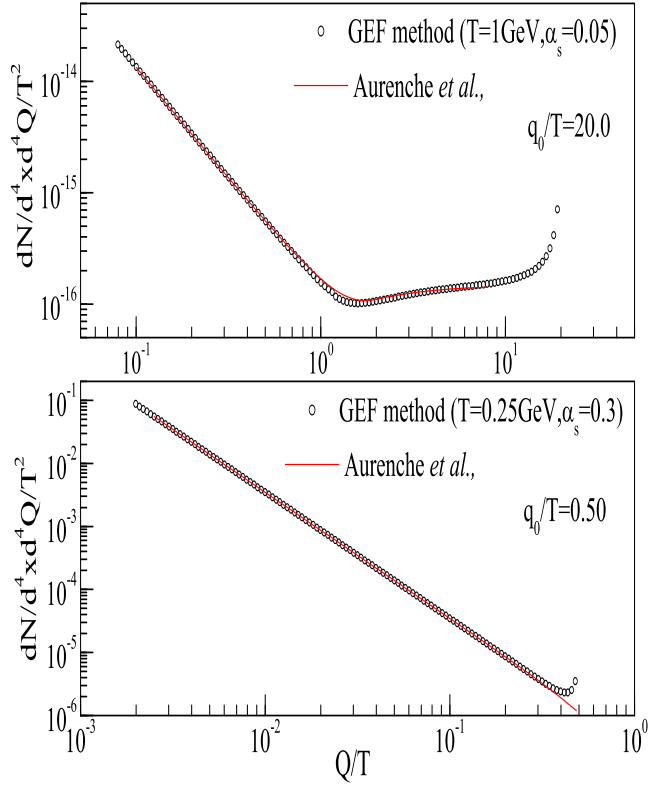}
\caption{  (a,b)~ Dilepton emission rates using GEF method shown in symbols and compared with results of \cite{liblpm} 
represented by blue lines. All the details are mentioned in figure labels and text. }
\label{gsnrates1}
 \end{figure}

\begin{figure}
\hspace{-1.05cm}
\includegraphics[height=16.cm,width=9.0cm]{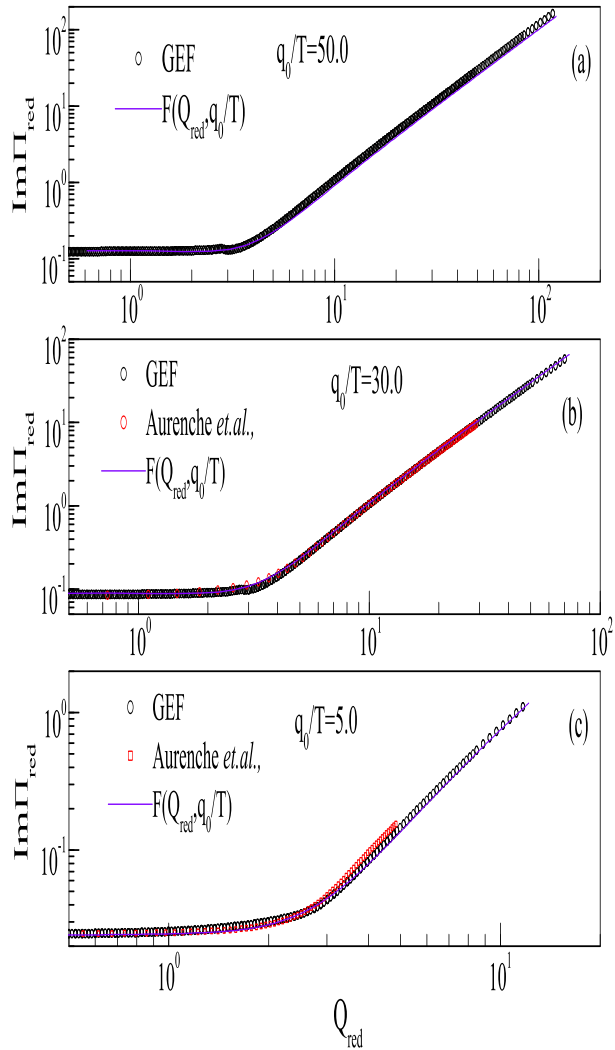}
\caption{$\Im \Pi_{red}$ plotted as a function of $Q_{red}=\frac{Q}{T}\sqrt{\frac{0.3}{\alpha_s}}$
 for various photon energies ($q_0/T$) mentioned in figure. The imaginary polarization tensor 
includes all contributions from transverse components
of bremsstrahlung, $\bf aws$, and also from the corresponding longitudinal parts.
The black circles represent the GEF method in Eq.\ref{impolargxmod}.
 The red circles represent the results of \cite{liblpm}.
The solid lines in violet color represent the results using  Eq.\ref{fq0qred}. } 
\label{impifitsh}
\end{figure}

\begin{figure}
\hspace{-1.5cm}
\includegraphics[height=16.cm,width=10.0cm]{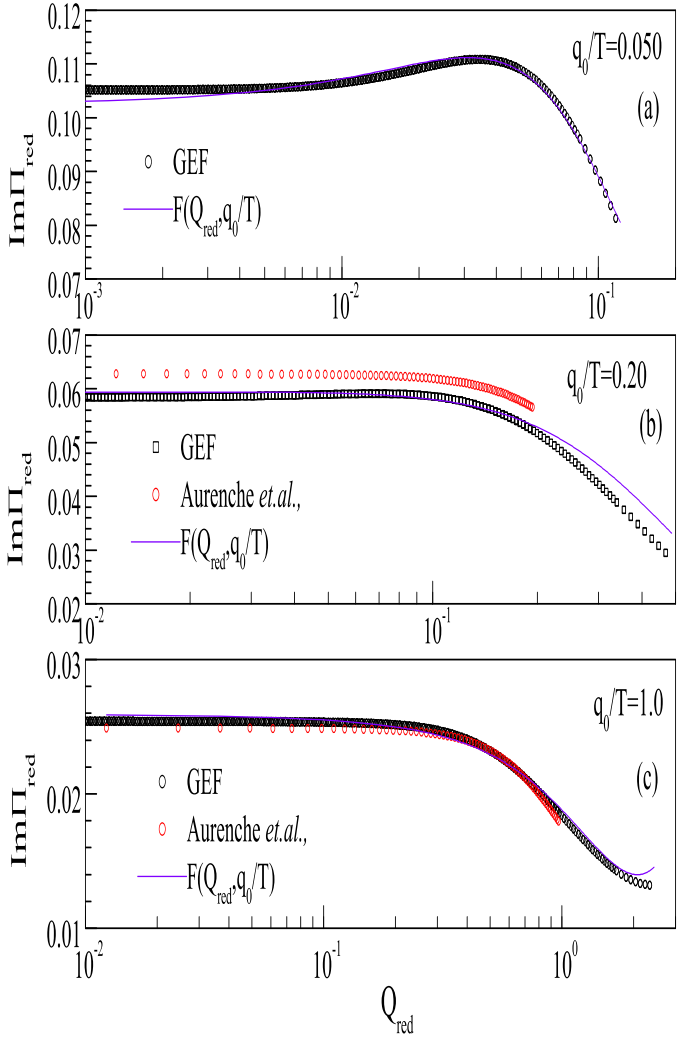}
\caption{$\Im \Pi_{red}$ plotted as a function of $Q_{red}$
 for low  $q_0/T$ values mentioned in figure. The details are as in previous figure \ref{impifitsh}.}
\label{impifitsl}
\end{figure}

 In the previous section, we used the results from the
iterations methods to obtain the $I^{b,a}_{T,L}$  values by integrating the ${\bf p_\perp}$distributions. 
We transformed these into  GEF  ($g^{b,a}_{T,L}$)  functions shown in Figs.\ref{btatgx}-\ref{blalgx}.
We fitted these by empirical functions given in  Eqs.\ref{btgxemp}-\ref{algxemp}.
Using the empirical functions, for any $p_0,q_0,Q^2,T,\alpha_s$  values,
 we can generate the  $I^{b,a}_{T,L}(x)$  values, 
 circumventing  the need to solve the integral equations. 
Thus, we have  empiricized the $I^{b,a}_{T,L}$  values 
in terms of GEF.  Hence,  using GEF and  the  $c^{b,a}_{T,L}$, 
the  imaginary part of photon retarded  polarization tensor   ($\Im \Pi_R$)  is calculated,  
as  in  Eq.\ref{impolargx} \cite{svsarxiv06}.
\begin{eqnarray}
\Im{\Pi^\mu}_{R\mu} &=& \frac{e^2N_c}{2\pi} \int_{-\infty}^\infty  dp_0 [n_F(r_0)-n_F(p_0)] \otimes \nonumber \\
&&\left(Tm_D^2\right) \left[ \frac{p_0^2+r_0^2}{2(p_0r_0)^2}\left(\frac{g^i_T \left(x_T\right)}{c^i_T}\right)+ \right. \nonumber \\
&& \left. \frac{1}{\sqrt{\left|p_0r_0\right|}}\frac{Q^2}{q^2} \left(\frac{1}{m_D}\right) \left(\frac{g^i_L \left(x_L\right)}{c^i_L}\right)  \right] 
\label{impolargx}
\end{eqnarray} 
 Here, the superscript $i$ denotes $ \{b,a\}$ depending on the value of the integration variable $p_0$.
\footnote{The factor T in (Tm$_D^2)$ in the Eq. \ref{impolargx} is arising from the  tilde transformation.
This extra T cancels the $\frac{m_D}{T}$ factor coming from tilde transformation of  $f,g$ functions. 
This T was missing in \cite{svsarxiv06}.}. 
We have calculated imaginary photon polarization tensor 
and dilepton emission rates using Eq.\ref{impolargx} and  made a detailed 
comparison with the results of  \cite{lpmdilep}. For this comparison, we generated reference 
results using the program provided by F. Gelis \cite{liblpm}.
The agreement of the GEF method of Eq.\ref{impolargx} with the results of  \cite{liblpm} was observed to be very good. 
As an example, we show the dilepton emission rates in Figure \ref{gsnrates1}. Figure
shows the GEF results in symbols compared with the results   of \cite{liblpm} (blue lines) 
at a photon ${q_0}$/T=20.0 and $\alpha_s$=0.05 (see (a) ). The GEF results were generated using T=1.0GeV.
Similarly in Figure (b) we show rates for
$q_0$/T= 0.50 and $\alpha_s$=0.30  (in fig.(b)). The GEF results were generated at T=0.25GeV. The agreement 
of GEF method with lines is seen to be  very good, except at the highest values of $\frac{Q}{T}$. This deviation is 
caused  because for the longitudinal partin Eq.\ref{impolargx},  we used photon momentum $\frac{Q^2}{q^2}$. When 
this is replaced with photon energy $\frac{Q^2}{q_0^2}$ as shown in Eq.\ref{impolargxmod}, 
the agreement of our results with \cite{liblpm} is very good in the full range of $Q/T$. In the remaining part of this paper, we 
use only Eq.\ref{impolargxmod}.
\begin{eqnarray}
\Im{\Pi^\mu}_{R\mu} &=& \frac{e^2N_c}{2\pi} \int_{-\infty}^\infty  dp_0 [n_F(r_0)-n_F(p_0)] \otimes \nonumber \\
&&\left(Tm_D^2\right) \left[ \frac{p_0^2+r_0^2}{2(p_0r_0)^2}\left(\frac{g^i_T \left(x_T\right)}{c^i_T}\right)+ \right. \nonumber \\
&& \left. \frac{1}{\sqrt{\left|p_0r_0\right|}}\frac{Q^2}{q_0^2} \left(\frac{1}{m_D}\right) \left(\frac{g^i_L \left(x_L\right)}{c^i_L}\right)  \right] 
\label{impolargxmod}
\end{eqnarray} 
\par
\begin{eqnarray}
Q_{red}&=&\frac{Q}{T}\sqrt{\frac{0.3}{\alpha_s}} \label{qred} \\
\Im{\Pi_{red}} &=& \frac{\Im{\Pi^\mu}_{R\mu}(Q^2,q_0,T,\alpha_s)}{T^2} \frac{0.30}{\alpha_s} \label{pired} 
\end{eqnarray}
We will present more results in a different way by defining reduced quantities.
After obtaining the $\Im{\Pi^\mu}_{R\mu}$ versus $Q^2,q_0,T,\alpha_s$ by using Eq.\ref{impolargxmod}, 
we define the reduced  polarization tensor as and reduced Q$_{red}$ as in Eqs.\ref{qred},\ref{pired}.
The reduced  polarization tensors are calculated for 
different photon energies, different coupling strengths and temperatures. We   
plotted these results in balck circles  in Figs.\ref{impifitsh},\ref{impifitsl} versus $Q_{red}$. 
For comparison,  results from \cite{liblpm} are shown in red symbols. The agreement of these two symbols is seen to be very good
from  low to very high  photon energies,  $q_0/T\sim 0.05-50.0$. 
\section{Phenomenology using Generalized Emission Functions}
\begin{figure}
\hspace{-0.85cm}
\includegraphics[height=12.cm,width=9.5cm]{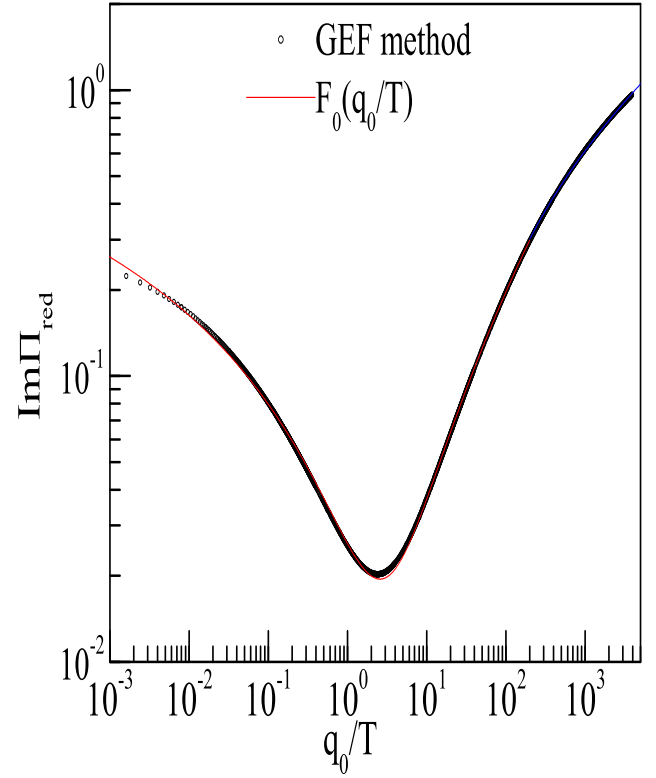}
\caption{  The reduced imaginary part of polarization tensor defined in Eq.\ref{pired}, versus $q_0/T$.
We have taken as $Q\sim10^{-4}q_0$. The solid curves are fits given in Eq.\ref{f0}. 
The symbols represent the results from GEF method using Eq.\ref{impolargxmod}.}
\label{snq0fit1}
\end{figure}

In this section, we obtain the phenomenological fits to virtual photon emission rates
from QGP. From the  Figures \ref{impifitsh},\ref{impifitsl}, it is clear that the reduced quantities depend on only two variables, {\it i.e.,}
instead of \{$Q^2,q_0,T,\alpha_s$\}, we need only \{$Q_{red},q_0/T$\} to generate $\Pi_{red}$ as in Eq.\ref{fdef}.
This observation was already reported in \cite{lpmdilep}.
\begin{equation}
\Im\Pi_{red}=F(Q_{red},\frac{q_0}{T}) \label{fdef}
\end{equation}
In the limit of $Q_{red}\rightarrow 0$, $F(Q_{red},\frac{q_0}{T}) \rightarrow F_0(\frac{q_0}{T}$).
To study this further,  we use  Eq.\ref{impolargxmod} to generate imaginary part of polarization tensor for various values of 
$Q^2,q_0,T,\alpha_s$. At  first, we generate the $Im\Pi$ at a very low $Q^2$,  $Q\sim10^{-4}q_0$, 
 for various values of $q_0/T,\alpha_s$.
Using the results, we construct $Q_{red},\Im\Pi_{red}$. The results are shown in Figure \ref{snq0fit1} by symbols
labeled GEF method.  The results for different $\alpha_s,T$  merge into  a single curve in Fig.\ref{snq0fit1}. 
We fitted this data by suitable functions 
as given in Eqs.\ref{f0} along with their parameters. 
These are two different fits, one for $q_0/T<$200 and the other  for $q_0/T>$100, with an overlap
between 100-200. These are represented by solid curves and labeled $F_0$ in figure. 
The $\Im\Pi_{red}$ below $q_0/T<0.020$ is approximately equal to this function $F_0$, as given by Eq.\ref{fredq0}.
However, it should be noted that at ultra soft photon energies, the present formalism needs corrections \cite{moore}.
\par
\begin{figure}
\hspace{-1.05cm}
\includegraphics[height=16.cm,width=9.250cm]{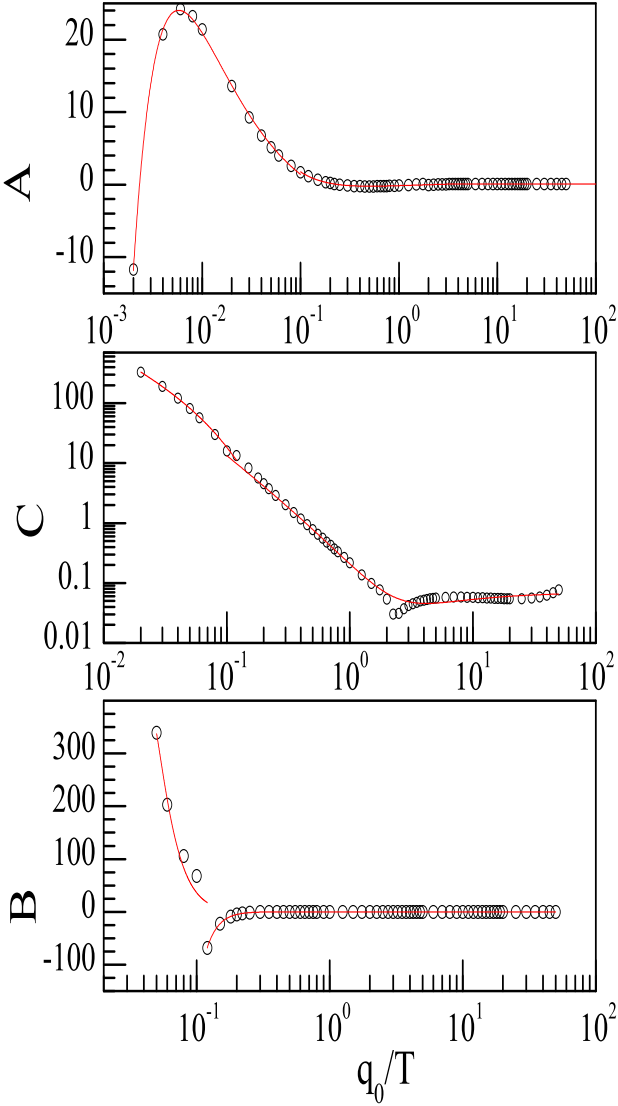}
\caption{  The A,B,C parameters versus $q_0/T$. The curves represent fits by suitable functional forms in 
Eqs.\ref{aparfit}, \ref{cparfit}, \ref{bparfit}  discussed in text. 
These are useful to generate $\Pi_{red}$ using Eq.\ref{fq0qred}. Apparently, at high $q_0/T$, 
these parameters are constant, however this is very misleading. The present fits generate quite well the 
small variations of these parameters over full region. 
Good quality  A,B,C fits are required because, the $Q^2$ plots are sensitive to these parameters
ane there is delicate cancellation of various terms in Eq.\ref{fq0qred}.} 
\label{abcparfit}
\end{figure}

For the case of finite $Q^2$, we made empirical fits by choosing a  function given in Eq.\ref{fq0qred}. In this function, A,B,C
parameters  are function of $q_0/T$ and are determined by fitting the   $Q^2$ plots for various $q_0/T$. 
These parameters values for various $q_0/T$  are tabulated and are shown in Figure \ref{abcparfit}. 
It is very important to have an empirical formula to generate A,B,C values.
Therefore, these A,C,B parameters were  fitted by different functional forms
as given in Eqs.\ref{aparfit},\ref{cparfit},\ref{bparfit}. The parameters are different for different
$q_0/T$ regions. Therefore, depending on the requirement, one can select the relevant parameter set to generate A,B,C values.
\footnote{It should be noted that the functional forms having difference
of power law  for these fits demand high precision of their parameters. Therefore, one should not truncate these parameters, especially 
the power exponents given by $p,p_1,p_2$.}
Using these formulae we get  A,B,C coefficients and we get   F$_0(x)$ from Eq.\ref{f0}.
We use these in Eq.\ref{fq0qred} to  generate $\Im\Pi_{red}$. These
phenomenological results are shown by solid curves in Figs.\ref{impifitsh},\ref{impifitsl}.
\begin{eqnarray}
F(Q_{red},x)&\approx&F_0(x)~~\mbox{for}~~x\le 0.020  \label{fredq0} \\
x&=&\frac{q_0}{T} \nonumber \\
F(Q_{red},x)&=&F_0(x)\frac{(1+A(x) Q_{red}+B(x)Q_{red}^4)}{(1+C(x) Q_{red}^2)} \label{fq0qred}
\end{eqnarray}
 \par
\appendix
\begin{eqnarray}
F_0(x)&=&a+b x^{p_1}+\frac{c}{\sqrt{x^{p_2}}} ~~\mbox{for}~x \le 200. \label{f0}\\
a &=& -2.99077 \nonumber \\
b &=& 0.0791399 \nonumber \\
c &=& 2.93755 \nonumber \\
p_1 &=& 0.371976 \nonumber \\
p_2 &=& 0.0288541 \nonumber \\
F_0(x)&=&a+b\sqrt{x^p}~~\mbox{for}~~x \ge 100.0 \\
a &=& -0.474129 \nonumber \\
b &=& 0.255163 \nonumber \\
p &=& 0.419646 \nonumber 
\end{eqnarray}

\begin{eqnarray}
A(x)&=&a+(b x^{p_1}-c x^{p_2}) ~~\mbox{for} ~~x \le 0.1 \label{aparfit} \\
a &=& -4.84554727516 \nonumber \\ 
b &=& 1.478772613744  \nonumber \\
c &=& 0.4963049612794 \nonumber \\
p_1 &=& -0.93213485133 \nonumber \\
p_2 &=& -1.1101191721  \nonumber \\
A(x)& & ~~\mbox{for}~0.1< x \le 3.5 \\
a &=& 0.33219586043 \nonumber \\
b &=& 1.34926189543   \nonumber \\
c &=& 1.82125018461   \nonumber \\
p_1 &=& -1.0422409717 \nonumber \\
p_2 &=& -0.86799168105  \nonumber \\
A(x)&=&A(3.5)~~\mbox{for} ~~x>3.50 
\end{eqnarray}
\begin{eqnarray}
C(x)&=&a+(b x^{p_1}-c x^{p_2}) ~~\mbox{for}~~x \le 0.15 \label{cparfit} \\
a &=& 7.48658946052   \nonumber \\ 
b &=& 13.1687711578  \nonumber \\
c &=& 19.9115694165  \nonumber \\
p_1& =& -1.0750505095 \nonumber \\
p_2 &=& -0.8514587339 \nonumber \\
C(x)& & ~~~~\mbox{for}~~x > 0.15 \\
a& =& 0.0692890486  \nonumber \\
b &=& 0.945340977  \nonumber \\  
c &=& 0.8057948943  \nonumber \\
p_1 &=& -1.4427187803  \nonumber \\
p_2 &=& -1.2054627582  \nonumber 
\end{eqnarray}

\begin{eqnarray}
B(x)&=&a+(b x^{p_1}-c x^{p_2}) ~~\mbox{for}~~x \le 0.20 \label{bparfit} \\
a &=& -0.132747  \nonumber \\
b &=& 0.0661859  \nonumber \\
c &=& 0.0642336  \nonumber \\
p_1&=& -4.76869  \nonumber \\
p_2&=& -4.77762   \nonumber \\
B(x)& & ~~\mbox{for} ~0.2<x \le 1.0 \\
a &=& 0.012385  \nonumber \\
b &=& 0.0924141  \nonumber \\
c &=& 0.0878252   \nonumber \\
p_1&=& -4.18536   \nonumber \\
p_2&=& -4.25575     \nonumber
\end{eqnarray}
\begin{eqnarray}
B(x)& & ~~\mbox{for} ~~~ x >1.0 \\
a &=& -0.0454930477  \nonumber \\
b &=& 0.1940150437   \nonumber \\
c &=& 0.1372510415    \nonumber \\
p_1& =& -0.237037465  \nonumber \\
p_2 &=& -0.42563026     \nonumber 
\end{eqnarray}
\par
\noindent
In conclusion, the  photon  emission  rates  from  the  quark  gluon plasma have been
studied as a function of photon mass, considering LPM  effects at various temperatures and strong coupling strengths.
 We defined generalized  dynamical variables $x_T,x_L$   for transverse and longitudinal components of bremsstrahlung
and  $\bf aws$ mechanism. In addition, we defined generalized emission functions (GEF)
namely $g^b_T(x_T)$,$g^a_T(x_T)$,$g^b_L(x_L)$,$g^a_L(x_L)$.
We  have obatined empirical fits to these GEF.
In terms of the  GEF,  we have  calculated the   imaginary part of retarded photon polarization tensor as a 
function of photon energy and mass, plasma temperature and strong coupling strengths.
 For  phenomenological applications,
we fitted the reduced imaginary polarization tensor by simple functions  and provided necessary parameters.\\
\acknowledgements{ I am thankful to my wife S.V. Ramalakshmi for co-operation during this work.}
\noindent


\begin{references}

\bibitem{landau1} L.D. Landau, I.Ya. Pomeranchuk, Dokl. Akad. Nauk. SSR {\bf 92}, 535 (1953)~;~{\it ibid.}~SSR {\bf 92}, 735 (1953).
\bibitem{migdal}A.B. Migdal, Phys. Rev. {\bf 103}, 1811 (1956).
\bibitem{arnold1}Peter  Arnold,  Guy  D.  Moore  and  Laurence  G.   Yaffe, JHEP {\bf 11} (2001) 057, [hep-ph/0109064].
\bibitem{arnold2}Peter  Arnold,  Guy  D.  Moore  and  Laurence  G.   Yaffe, JHEP {\bf 12} (2001) 009, [hep-ph/0111107]
\bibitem{svsprc}S. V. S. Sastry, Phys. Rev. {\bf C67}, 041901(R) (2003), [hep-ph/0211075] ; [hep-ph/0208103]
\bibitem{alther}T. Altherr, P.V. Ruuskanen, Nucl. Phys. {\bf B380}, 377 (1992).
\bibitem{thoma}M.H. Thoma, C.T. Traxler, Phys. Rev. {\bf D56}, 198 (1997), [hep-ph/09701354]
\bibitem{lpmdilep}P. Aurenche, F. Gelis, Guy D. Moore and  H.  Zaraket, JHEP {\bf 12} (2002) 006, [hep-ph/0211036].
\bibitem{braaten}E. Braaten, R.D. Pisarski, Nucl. Phys. {\bf B337}, 569 (1990).
\bibitem{svsarxiv06}S.V. Suryanarayana, Phys. Rev. {\bf C75},021902(R) (2007),[hep-ph/0606056]
\bibitem{svsvar}S.V. Suryanarayana,  hep-ph/0609096, work in progress.
\bibitem{liblpm}F. Gelis, libLPM-v1, \\ http://www-spht.cea.fr/articles/T02/150/libLPM/
\bibitem{moore}Guy D. Moore and Jean-Marie Robert, arXiv:hep-ph/0607172.
\end{references}
\end{document}